\documentclass[aps,prl,twocolumn,superscriptaddress]{revtex4-2}
\usepackage{graphicx}
\usepackage{mathrsfs}
\usepackage{bm}
\usepackage{amsmath}
\usepackage{color}
\usepackage{dcolumn}
\usepackage{epstopdf}
\usepackage{dsfont}
\usepackage{amssymb}
\usepackage{tabularx}
\usepackage{array,ulem}
\usepackage{float}
\usepackage{colordvi}
\usepackage{verbatim}
\usepackage{hyperref}

\hypersetup{
colorlinks=true,
linkcolor=blue,
filecolor=blue,
urlcolor=blue,
citecolor=blue,
}

\begin{document}
\title{Tunable Two-Dimensional Dirac-Weyl Semimetal Phase Induced by Altermagnetism}

\author{Lizhou Liu}
\affiliation{College of Physics, Hebei Normal University, Shijiazhuang 050024, China}
\affiliation{International Center for Quantum Materials, School of Physics, Peking University, Beijing 100871, China}

\author{Qing-Feng Sun}
\email[Correspondence author:~~]{sunqf@pku.edu.cn}
\affiliation{International Center for Quantum Materials, School of Physics, Peking University, Beijing 100871, China}
\affiliation{Hefei National Laboratory, Hefei 230088, China}

\author{Ying-Tao Zhang}
\email[Correspondence author:~~]{zhangyt@mail.hebtu.edu.cn}
\affiliation{College of Physics, Hebei Normal University, Shijiazhuang 050024, China}

\date{\today}

\begin{abstract}
We demonstrate a tunable Dirac-Weyl semimetal phase in two dimensions, realized by introducing in-plane $d$-wave altermagnetism into a Dirac system. This phase hosts both a central Dirac point and momentum-separated Weyl points connected by Fermi line edge states. The Weyl point positions—and thus the edge-state connectivity—can be continuously tuned by rotating the altermagnetic axis. In contrast, out-of-plane altermagnetism gaps part of the bulk spectrum while preserving a single Dirac point accompanied by chiral edge modes, as evidenced by quantized edge polarization.
Our findings provide a tunable platform for manipulating Dirac-Weyl physics and topological edge transport in two dimensions.
\end{abstract}

\maketitle

\textit{Introduction---}Topological semimetals exhibit unconventional band structures and symmetry-protected boundary modes, with Dirac and Weyl semimetals as paradigmatic examples~\cite{Armitage2018, Lv2021}. Weyl semimetals feature twofold-degenerate band crossings—Weyl points—that act as Berry curvature monopoles and require broken inversion or time-reversal symmetry~\cite{Wan2011, Weng2015, Soluyanov2015, Vazifeh2013, Yan2016, Fang2012, Kaushik2025, Forslund2025}, giving rise to phenomena such as the chiral anomaly and surface Fermi arcs~\cite{Nielsen1983, Son2013, Huang2015, Li2016, Xu2015, Lv2015}. In contrast, Dirac semimetals host fourfold-degenerate crossings stabilized by time-reversal, inversion, and crystal symmetries~\cite{Young2012, Wang2012, Liu2014a, Liu2014b, Wang2013, Liu2024, Borisenko2014, Sanchez-Barriga2023, Wieder2016, Yang2022}, and exhibit ultrahigh mobility and giant magnetoresistance~\cite{Liang2015}.

 While Dirac and Weyl fermions were traditionally viewed as mutually exclusive due to incompatible symmetry requirements, recent studies have demonstrated their coexistence in specially engineered systems, giving rise to Dirac-Weyl semimetals~\cite{Gao2018, Wu2024, Long2023, Jin2024}. 
Such novel quantum states facilitate a direct interplay between high-mobility Dirac carriers and topologically nontrivial Weyl nodes, simultaneously supporting coherent bulk transport and symmetry-controllable edge-state configurations.
This dual character enables symmetry-controlled redistribution of Berry curvature, programmable Fermi arc connectivity, and momentum-selective topological transport--hallmark features that go beyond the scope of conventional semimetallic behavior.
However, most existing proposals rely on nonsymmorphic space groups or artificial photonic analogs, and lack both tunability and electronic realizability.
The realization of an intrinsically electronic and magnetically controllable Dirac-Weyl semimetal would constitute a significant step toward reconfigurable topological phases and device-relevant quantum functionalities.

In this Letter, we propose a symmetry-guided strategy to realize a tunable Dirac-Weyl semimetal phase in a two-dimensional magnetic system. By introducing in-plane $d$-wave altermagnetism into a Dirac semimetal, we induce a topological phase hosting both a central Dirac point and momentum-separated Weyl points, with the latter connected by Fermi line edge states. Crucially, the Weyl point positions, and hence the edge-state connectivity, can be continuously tuned by rotating the in-plane altermagnetic orientation. In contrast, switching the magnetic order to the out-of-plane direction partially gaps the bulk spectrum while preserving a single Dirac point, and simultaneously generates chiral edge modes protected by topology. These results establish altermagnetic order as a versatile control knob for engineering symmetry-tunable Dirac-Weyl phases in realistic two-dimensional electronic systems.

\begin{figure}
  \centering
  \includegraphics[width=7.8 cm,angle=0]{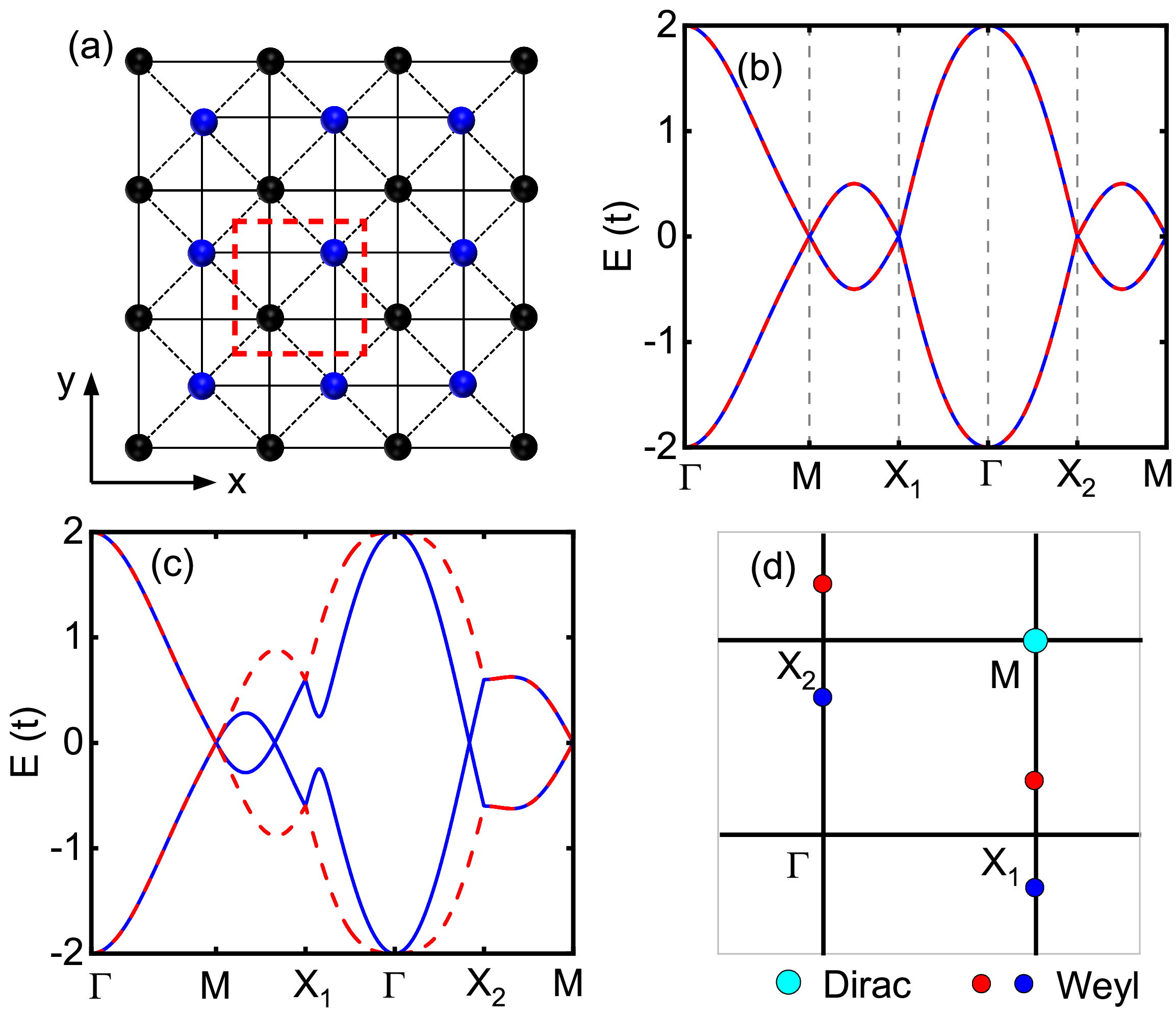}
  \caption{(a) Schematic illustration of the $\sqrt{2} \times \sqrt{2}$ square lattice.
The unit cell is consist of two sublattices in black and blue as indicated inside the dashed red rectangle.
The dashed lines and black solid lines represent the nearest-neighbor and second-nearest-neighbor couplings, respectively.
(b) Bulk band structure along high-symmetry lines in the absence of altermagnetic order, showing Dirac points at $X_1$, $X_2$, and $M$ points.
(c) Bulk band structure under $x$-direction $d$-wave altermagnetism, where the Dirac points at $X_1$ and $X_2$ split into pairs of Weyl points.
Bulk bands are plotted using blue solid and red dashed lines to highlight their degeneracy.
(d) Schematic distribution of Dirac and Weyl points in the Brillouin zone.
The parameters used are $t = -1$, $t_2 = 0$, $t_s= 0.5$, and $J = 0.3$.
  }
  \label{fig1}
\end{figure}

\textit{System Hamiltonian---} We consider a two-dimensional square lattice composed of two sublattices, as illustrated in Fig.~\ref{fig1}(a). 
The two sublattices are displaced alternately along the $z$ direction, forming a buckled configuration that preserves nonsymmorphic symmetries, enabling the emergence of symmetry-protected Dirac points in the presence of spin-orbit coupling~\cite{Young2015}. 
The tight-binding Hamiltonian in momentum space is given by
\begin{equation}
\begin{aligned}
H(\mathbf{k}) &= 2t \cos\left(\frac{k_x}{2}\right) \cos\left(\frac{k_y}{2}\right) \tau_x + t_2 (\cos k_x + \cos k_y) \\
&\quad + t_s \tau_z (\sigma_y \sin k_x - \sigma_x \sin k_y),
\end{aligned}
\end{equation}
where $\tau_{x,z}$ and $\sigma_{x,y}$ are Pauli matrices acting on sublattice and spin degrees of freedom, respectively. The first term describes nearest-neighbor hopping between the two sublattices, while the second captures second-neighbor hopping within each sublattice. The third term introduces spin-orbit coupling via next-nearest-neighbor interactions, characteristic of nonsymmorphic systems, and leads to spin-momentum locking.

To preserve particle-hole symmetry and align the Dirac points at $X_1$, $X_2$, and $M$ to a common energy, we set $t_2=0$, which enforces $\{H,\tau_y\}=0$, unless otherwise specified, following Ref.~\cite{Young2015}. 
We alos note the a finite $t_2$ breaks particle-hole symmetry and shifts the $M$-point Dirac node away from zero energy, but it does not affect the robustness of the Dirac-Weyl semimetal phase, as discussed below.
This creates a symmetry-protected platform to explore topological band crossings. To realize a Dirac-Weyl semimetal phase, we introduce a \( d \)-wave altermagnetic exchange field. Altermagnetism is a collinear magnetic order that breaks time-reversal symmetry \( \mathcal{T} \) while preserving inversion \( \mathcal{P} \) and fourfold rotation \( \mathcal{C}_4 \), resulting in momentum-dependent spin splitting without net magnetization~\cite{Smejkal2022, Smejkal2022a, Mazin2024, McClarty2024, Gu2025}. Unlike a uniform Zeeman field, it produces anisotropic spin textures that selectively lift Dirac degeneracies and generate Weyl nodes. Its symmetry-selective nature and topological versatility~\cite{Li2024, Li2025, Fernandes2024, Liu2025, Cheng2024, Cheng2} make it an effective tool for engineering tunable nodal structures.

Guided by these insights, we introduce an in-plane \( d \)-wave altermagnetic texture modeled by
\begin{equation}
H_{\mathrm{AM}} = J (\cos k_x - \cos k_y) \tau_0 \sigma_x,
\label{eq:Ham}
\end{equation}
where \( J \) is the exchange coupling strength, \( \tau_0 \) is the identity in sublattice space, and \( \sigma_x \) denotes spin polarization along the \( x \)-direction. This term vanishes along the nodal lines \( k_x = \pm k_y \) and changes sign across them, resulting in anisotropic spin splitting across the Brillouin zone \cite{addr5,addr6}. It acts as a momentum-selective perturbation that preserves inversion symmetry \( \mathcal{P} \) and fourfold rotation symmetry \( \mathcal{C}_4 \), while breaking mirror symmetries \( \mathcal{M}_x \) and \( \mathcal{M}_y \)~\cite{Smejkal2022a}, in close analogy to the symmetry structure established in minimal altermagnetic models~\cite{Roig2024}. As we demonstrate below, this controlled symmetry breaking lifts specific Dirac degeneracies and drives the emergence of Weyl nodes.
The $d$-wave altermagnetic order is essential here, since its momentum dependence uniquely vanishes at $M$ but remains finite at $X_{1,2}$, thereby enabling the selective Dirac-Weyl phase.

\textit{Dirac-Weyl Semimetal and Symmetry Breaking---}
We begin with the two-dimensional Dirac semimetal proposed by Young and Kane~\cite{Young2015}, which hosts three symmetry-protected Dirac points at \( X_1 = (\pi, 0) \), \( X_2 = (0, \pi) \), and \( M = (\pi, \pi) \). These fourfold-degenerate nodes arise from nonsymmorphic lattice symmetries and spin-orbit coupling. In the absence of magnetism, the system respects time-reversal symmetry (\( \mathcal{T} \)), inversion symmetry (\( \mathcal{P} \)), and two glide mirror symmetries (\( \mathcal{G}_x \), \( \mathcal{G}_y \)) associated with the \( \sqrt{2} \times \sqrt{2} \) distortion. Specifically, the Dirac points at \( X_1 \) and \( X_2 \) are protected by the anti-commutation of \( \mathcal{T} \) with the corresponding glide operations~\cite{Young2015}. The band structure under these symmetries [Fig.~\ref{fig1}(b)] reproduces the original results~\cite{Young2015}, and low-energy \( \mathbf{k} \cdot \mathbf{p} \) expansions are provided in Sec. II of the Supplemental Material (SM)\cite{SM2025}.

To explore magnetic symmetry breaking, we introduce an in-plane \( d \)-wave altermagnetic exchange field of the form \( H_{\mathrm{AM}} = J (\cos k_x - \cos k_y)\, \tau_0 \sigma_x \). This perturbation selectively lifts the fourfold degeneracies at \( X_1 \) and \( X_2 \), each splitting into a pair of twofold-degenerate Weyl nodes, while leaving the Dirac point at \( M \) intact due to its location along the nodal lines of the altermagnetic texture and its symmetry protection by \( \mathcal{P} \) and \( \mathcal{C}_4 \) [Fig.~\ref{fig1}(c)]. 
The stability of the Dirac node at \( M \)  relies crucially on inversion symmetry. Once this symmetry is broken, for instance by a staggered sublattice potential, the \( M \)-point Dirac node is destroyed.
The resulting semimetallic state features coexisting Dirac and Weyl nodes, which we term a \textit{Dirac-Weyl semimetal} [Fig.~\ref{fig1}(d)].

This transition is captured by a \( \mathbf{k} \cdot \mathbf{p} \) expansion near \( X_1 = (\pi, 0) \), yielding the effective Hamiltonian~\cite{Young2015}
\begin{equation}
H^{\text{eff}}_{X_1} = (-t\, \tau_x - t_s \, \tau_z \sigma_y)\, q_x - t_s \, \tau_z \sigma_x\, q_y - 2J\, \tau_0 \sigma_x,
\end{equation}
where \( \mathbf{q} = (q_x, q_y)\) denotes momentum relative to \( X_1 \). 
The altermagnetic term \( -2J \tau_0 \sigma_x \) breaks time-reversal symmetry \( \mathcal{T} \), leading to a pair of Weyl nodes located at \( q_y = \pm 2J / t_s \), as detailed in Sec.~II of the SM~\cite{SM2025}.
Therefore, the separation between the Weyl nodes increases with the exchange strength \( J \), as confirmed numerically in Fig.~S1 of the SM~\cite{SM2025}, which shows the evolution of the bulk band structure, Weyl node positions, and edge spectra as \( J \) increases.

A similar analysis at \( X_2 = (0, \pi) \) reveals a Weyl node pair rotated by 90°, while the Dirac point at \( M \) remains gapless. To further characterize the emergent Weyl nodes, we project the Hamiltonian onto the \( \tau_z = -1 \) subspace, yielding a two-band model:
\begin{equation}
H_{\text{Weyl}}(\mathbf{q}) = t_s \sigma_y q_x + t_s \sigma_x q_y,
\end{equation}
which describes a conventional two-dimensional Weyl cone. The two nodes carry opposite topological charges \( \chi = \pm 1 \), confirmed via Berry curvature integration (see Sec.~IV of the SM~\cite{SM2025}). These results establish the Dirac-Weyl semimetal as a tunable topological phase driven by symmetry-selective magnetic ordering.

\begin{figure}
  \centering
  \includegraphics[width=8.2 cm,angle=0]{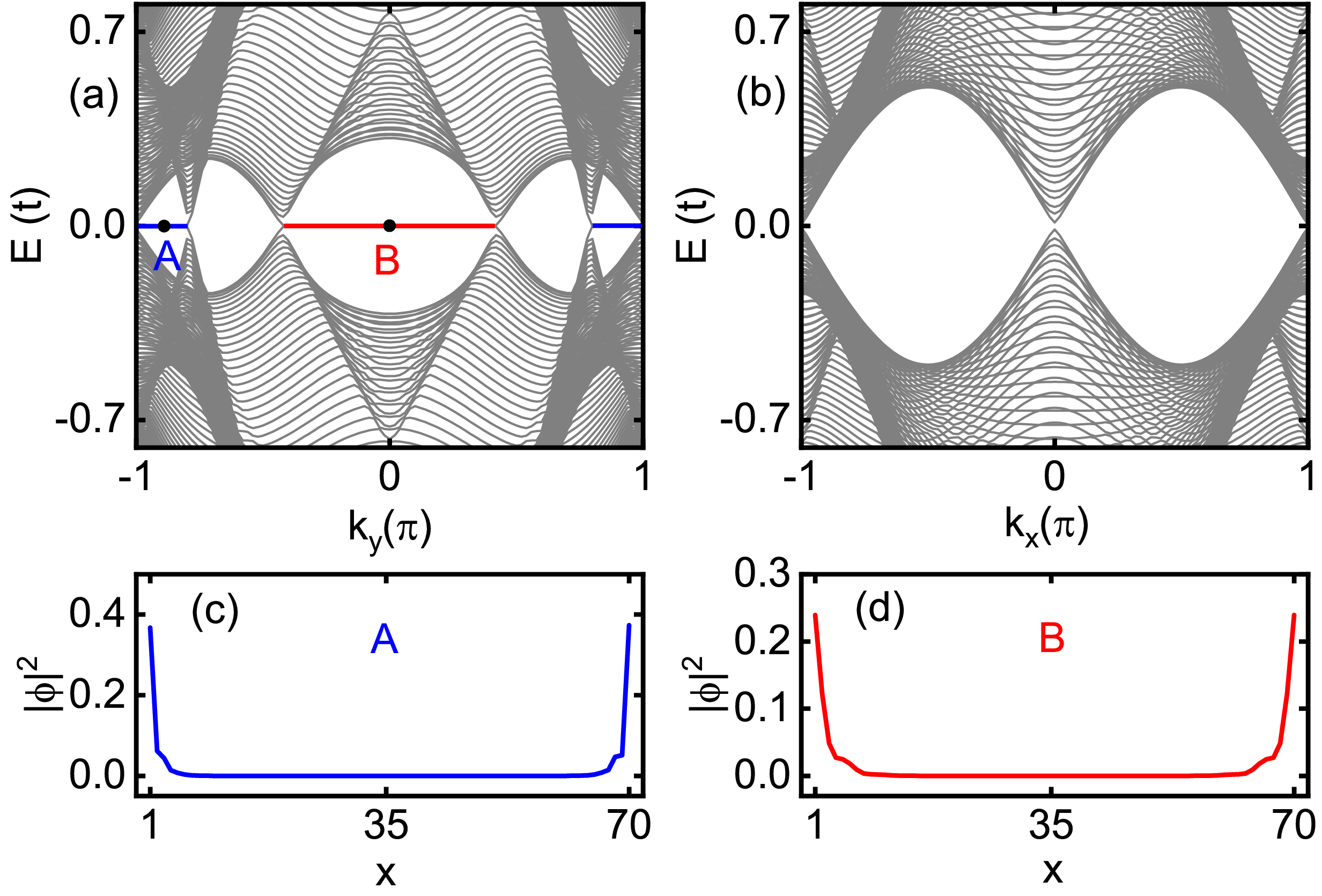}
  \caption{(a) Energy spectrum as a function of $k_y$ with open boundary conditions along the $x$ direction, showing Fermi line edge states connecting Weyl points. The two edge states highlighted by labels A (blue) and B (red), respectively.
(b) Energy spectrum as a function of $k_x$ under open boundary conditions along $y$, where no Fermi line states are observed.
(c), (d) Real-space probability distributions $|\phi|^2$ along $x$ for the edge states A and B marked in (a).
The parameters used are $t = -1$, $t_2 = 0$, $t_s = 0.5$, and $J = 0.3$.}
  \label{fig2}
\end{figure}

\textit{Fermi line States and Edge Localization---}
 A hallmark of the three-dimensional Weyl semimetal phase is 
the emergence of Fermi arc states that connect Weyl nodes of opposite chirality. 
For the two-dimensional Weyl semimetal phase, two Fermi point states typically emerge 
[see Figs. S2(b) and S2(d) of the SM~\cite{SM2025}]. 
When considering the entire energy range, these two Fermi point states extend into a line (which we refer to as the Fermi line) connecting the two Weyl nodes.
As shown in Fig.~\ref{fig2}(a), the boundary spectrum under open boundary conditions along the $x$ direction exhibits two pairs of nearly flat edge modes, localized near the $X_1$ and $X_2$ valleys. These edge states arise from the nontrivial bulk topology: each Weyl node acts as a source or sink of Berry curvature, and their projections onto the surface Brillouin zone are connected by gapless boundary modes. The in-plane $d$-wave altermagnetic field induces a momentum-space separation of Weyl nodes along $k_y$, keeping their surface projections distinct along the $x$ edge and thus enabling the formation of topologically protected Fermi line.
The evolution of these Fermi line states with increasing \( J \) is presented in Fig.~S1(b) of the SM~\cite{SM2025}, which shows that the momentum separation between edge modes grows as the Weyl nodes move apart.

By contrast, the edge spectrum along the $y$ direction [Fig.~\ref{fig2}(b)] shows no gapless edge states. This is because the Weyl nodes project onto the same momentum along $k_x$, leading to vanishing net chirality and precluding Fermi line formation. This directional anisotropy highlights that a nonzero projected topological charge is essential for realizing boundary-localized modes in two-dimensional Weyl systems.

The edge localization of these Fermi line modes is confirmed by real-space wavefunction profiles [Figs.~\ref{fig2}(c), \ref{fig2}(d)], which show strong amplitude near $x$-oriented edges. In contrast, the Dirac node at $M$ does not contribute to edge states, reflecting its symmetry-protected origin rather than topological charge. These findings collectively demonstrate the coexistence of bulk Dirac and Weyl nodes and establish the presence of one-dimensional topological boundary modes controlled by momentum-selective symmetry breaking.
This coexistence not only realizes a long-sought hybrid topological state but also enables unprecedented functionalities: Dirac fermions contribute ultrafast carriers with linear dispersion, while Weyl nodes introduce momentum-selective Berry curvature sources, enabling directional topological transport and anomalous Hall responses. Moreover, the engineered momentum-space separation allows the emergence of tunable Fermi line states, whose anisotropic edge connectivity may serve as a fingerprint for designer topological circuitry.

Since in realistic materials the second-neighbor hopping $t_2$ is typically much smaller than the nearest-neighbor hopping, the approximation $t_2=0$ adopted in our minimal model is physically reasonable. Nevertheless,
to further examine the stability of the Dirac-Weyl semimetal phase, we introduce a second-neighbor hopping term \( t_2 \), which breaks particle-hole symmetry by violating \( \{H, \tau_y\} = 0 \). 
As shown in Fig.~S2 of the SM~\cite{SM2025}, finite \( t_2 \) shifts the Dirac point at \( M \) away from zero energy, yet the Weyl node splitting at \( X_1 \) and \( X_2 \) remains intact. 
The associated Fermi line edge states are also preserved, demonstrating that the Dirac-Weyl phase is robust against such symmetry-breaking perturbations.

\begin{figure}
  \centering
  \includegraphics[width=7.2 cm,angle=0]{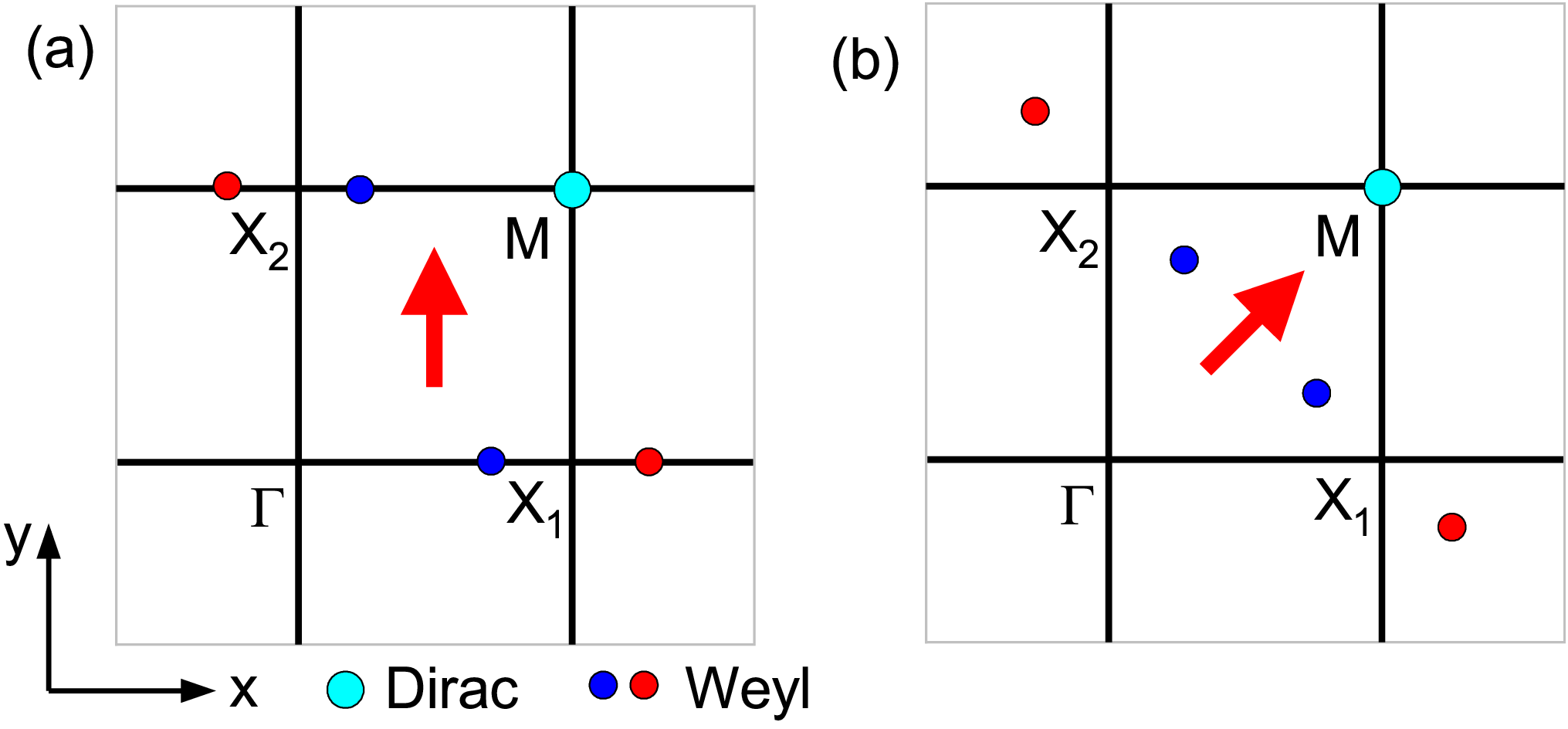}
  \caption{
Momentum-space distributions of Dirac and Weyl points in the Brillouin zone under different orientations of the $d$-wave altermagnetic order.
(a) For an altermagnetic field aligned along the $y$ direction, pairs of Weyl points (red and blue dots) emerge near $X_1$ and $X_2$ and split along the $x$ direction, while the Dirac point (cyan) at $M$ remains intact.
(b) Rotating the altermagnetic direction causes the Weyl points to shift accordingly, with their splitting direction always perpendicular to the altermagnetic orientation.
The red arrow indicates the direction of the in-plane altermagnetism.
}
  \label{fig3}
\end{figure}

\textit{Directional control of Weyl nodes---}
A salient feature of the Dirac-Weyl semimetal phase is the directional tunability of Weyl node positions via the orientation of the in-plane $d$-wave altermagnetic order. As illustrated in Fig.~\ref{fig3}(a), when the altermagnetic axis $\hat{\mathbf{m}}$ is aligned along $\hat{y}$, the Dirac points at $X_1$ and $X_2$ split into pairs of Weyl nodes displaced along the $k_x$ direction. Upon rotating $\hat{\mathbf{m}}$, the momentum-space separation of Weyl points rotates accordingly [Fig.~\ref{fig3}(b)].

Based on the effective Hamiltonian analysis (see Sec.~II of the SM~\cite{SM2025}), the Weyl nodes induced near \( X_1 \) and \( X_2 \) always split along the direction perpendicular to the altermagnetic vector $\hat{\mathbf{m}}$.
The altermagnetic term $-2J \tau_0 \boldsymbol{\sigma} \cdot \hat{\mathbf{m}}$ breaks time-reversal symmetry $\mathcal{T}$ and lifts the Dirac-point degeneracy by not commuting with the spin-orbit coupled Dirac Hamiltonian, except along momenta parallel to $\hat{\mathbf{m}}$. As a result, Weyl nodes emerge along the direction perpendicular to $\hat{\mathbf{m}}$, and rotating $\hat{\mathbf{m}}$ reorients their separation axis while preserving topological charge.
This mechanism provides a symmetry-respecting platform for controlling both nodal topology and Fermi line connectivity through the in-plane magnetization direction.

\begin{figure}
  \centering
  \includegraphics[width=7.8 cm,angle=0]{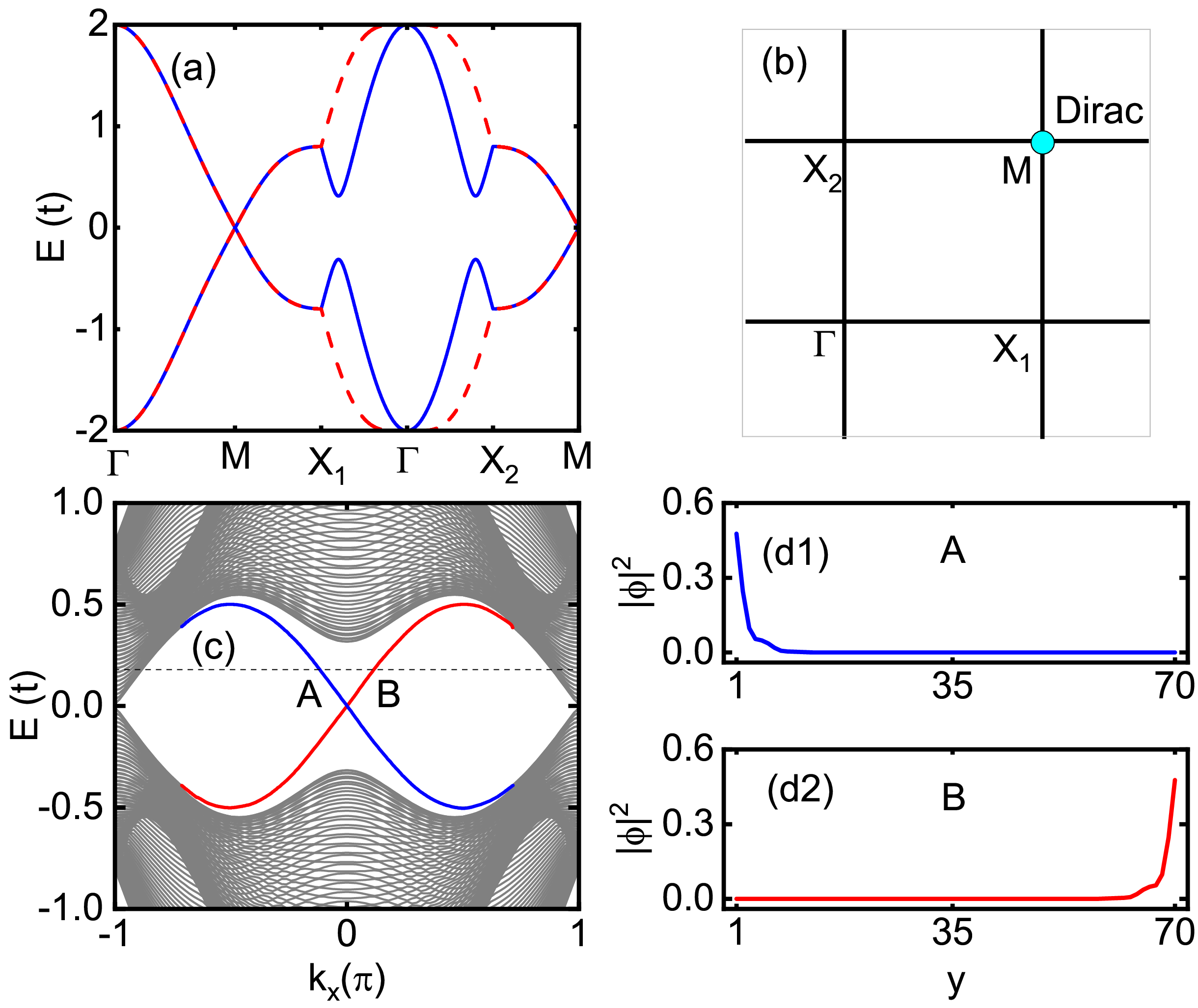}
  \caption{(a) Bulk band structure along high-symmetry lines under out-of-plane $d$-wave altermagnetic order, showing that only the Dirac point at $M$ remains gapless.
(b) Schematic distribution of the residual Dirac point in the Brillouin zone.
(c) Energy spectrum as a function of $k_x$ under open boundary conditions along $y$, revealing a pair of chiral edge states (labeled A and B).
(d1), (d2) Probability distributions $|\phi|^2$ along the $y$ direction for the edge states A and B, respectively, confirming their localization at opposite boundaries.
The parameters used are $t = -1$, $t_2 = 0$, $t_s= 0.5$, and $J = 0.4$.
  }
  \label{fig4}
\end{figure}

\textit{Chiral edge states in a Dirac semimetal---}
Rotating the $d$-wave altermagnetic field out of the plane fundamentally alters the nodal structure of the Dirac-Weyl semimetal. 
The out-of-plane altermagnetic field $(H_{\mathrm{AM}} = J (\cos k_x - \cos k_y) \tau_0 \sigma_z)$ introduces coupling between previously degenerate bands at \( X_1 \) and \( X_2 \), leading to hybridization and the opening of a full energy gap [Fig.~\ref{fig4}(a)]. This gap emerges due to symmetry breaking and band mixing induced by the magnetic perturbation.
In contrast, the Dirac point at $M$ remains gapless, protected by inversion and $ \mathcal{C}_4$ rotational symmetries, with only high order corrections induced by the out-of-plane field [Fig.~\ref{fig4}(b)]. These results are corroborated by symmetry-resolved \( \mathbf{k} \cdot \mathbf{p} \)  expansions, presented in Sec.~II of the SM~\cite{SM2025}.

Although the bulk remains gapless, a pair of chiral edge modes emerges [Fig.~\ref{fig4}(c)], revealing a topological character beyond that of conventional Chern insulators. Real-space wave-function profiles [Figs.~\ref{fig4}(d1), \ref{fig4}(d2)] confirm that these modes are sharply localized at opposite edges, consistent with unidirectional boundary propagation. 
To further establish their topological origin, we compute the momentum-resolved wave polarizations $P_y(k_x{=}0)$ and $P_x(k_y{=}0)$ along high-symmetry lines using the Berry connection formalism~\cite{Liu2017,Xie2021}. 
As shown in Sec.~IV of the SM~\cite{SM2025}, both polarizations exhibit quantized values of $\pm 1/2$, protected by crystalline symmetries. 
These quantized dipole moments serve as bulk topological invariants in the gapless phase and confirm the existence of a single chiral mode per edge. 
In this state, the symmetry-protected Dirac point at $M$ coexists with quantized chiral edge states. 
These edge states originate from the gapped Dirac nodes at $X_{1,2}$, providing a novel mechanism for realizing boundary modes without a full bulk gap.
This coexistence offers a promising platform for robust signal propagation in semi-metallic regimes.

\textit{Conclusion---}
We have proposed a tunable two-dimensional Dirac-Weyl semimetal phase induced by $d$-wave altermagnetic order in a nonsymmorphic Dirac semimetal. The in-plane altermagnetic texture selectively splits Dirac points into Weyl nodes, enabling Fermi line edge states and directional control of their connectivity without breaking crystalline symmetries. 
The distance between the Weyl nodes increases with the strength of the altermagnetic exchange field. And, the direction of Weyl node separation—and hence the Fermi line orientation in momentum space---can be tuned by rotating the altermagnetic axis.
Out-of-plane altermagnetism gaps the Weyl nodes while preserving a symmetry-protected Dirac point, yielding a gapless phase with quantized chiral edge modes. 
For comparison, Ref.~\cite{Jin2024} realized a two-dimensional Dirac-Weyl phase through strain engineering, where tunability arises from modifying the Fermi velocities of Dirac and Weyl cones. 
In contrast, our work is based on a momentum-dependent $d$-wave altermagnetic order, which provides an intrinsically electronic and symmetry-guided mechanism. 
This approach allows continuous control of Weyl-node positions and edge-state connectivity by rotating the altermagnetic axis, in sharp contrast to the discrete strain configurations required in Ref.~\cite{Jin2024}.

Our findings open a new avenue for symmetry-controlled topological states with coexisting Dirac and Weyl fermions in minimal two-dimensional  systems. The demonstrated directional control of Weyl nodes, tunable Fermi line, and chiral edge modes offers rich possibilities for momentum-space interferometry, topological signal routing, and low-dissipation edge transport. 
Recent experiments have confirmed d-wave--like altermagnetic order in MnTe thin films~\cite{Amin2024}, while similar d-wave spin textures have been identified in RuO$_2$~\cite{Lin2024}.
In parallel, several material predictions of nonsymmorphic Dirac semimetals have been reported in two-dimensional systems, such as monolayer HfGeTe and chemically modified group-VA monolayers~\cite{Guan2017,Jin2019}.
The growing experimental accessibility of altermagnetism~\cite{Feng2022,Bose2022,Karube2022,Bai2022} and the recent material predictions of nonsymmorphic Dirac semimetals~\cite{Guan2017,Jin2019} demonstrate that the essential symmetry conditions required by our model can be realized in real materials.
Hybrid structures combining d-wave--like altermagnets (e.g., MnTe or RuO$_2$) with nonsymmorphic Dirac semimetals thus provide promising platforms for realizing the tunable Dirac-Weyl phase proposed in our model.
Importantly, the Dirac-Weyl semimetal phase proposed here is driven entirely by the intrinsic d-wave spin splitting induced by altermagnetism rather than by fine-tuned parameters, implying that our mechanism remains valid whenever the required crystalline and magnetic symmetries are present.
These insights also provide clear guidance for future first-principles or material-design efforts aimed at realizing the proposed Dirac-Weyl semimetal phase in real systems.

This work was financially supported by the National Natural Science Foundation of China (Grants No. 12074097, No. 12374034, and No. 11921005),
Natural Science Foundation of Hebei Province (Grant No. A2024205025),
the National Key R and D Program of China (Grant No. 2024YFA1409002),
and the Innovation Program for Quantum Science and Technology (Grant No. 2021ZD0302403).

\end{document}